\documentclass[12pt]{emulateapj} 
\usepackage{amsmath}
\usepackage{amssymb}
\usepackage{amstext}
\usepackage{apjfonts}
\usepackage{graphicx}
\usepackage{epstopdf}
\usepackage{epsfig}
\usepackage{color}
\definecolor{myColor}{rgb}{0.9,0.9,0.9}    
\begin{document}
\renewcommand\bottomfraction{.9}
\shorttitle{Trojan Companions to Transiting Exoplanets}
\title{Empirical Constraints on Trojan Companions and Orbital Eccentricities in 25
Transiting Exoplanetary Systems}
\author{N.\ Madhusudhan\altaffilmark{1} \& Joshua N.\ Winn\altaffilmark{1}}
\altaffiltext{1}{Department of Physics and Kavli Institute for 
Astrophysics and Space Research, MIT, Cambridge, MA 02139; {\tt 
nmadhu@mit.edu}}
\begin{abstract}

  We present a search for Trojan companions to 25 transiting
  exoplanets. We use the technique of Ford \& Gaudi, in which a
  difference is sought between the observed transit time and the
  transit time that is calculated by fitting a two-body Keplerian
  orbit to the radial-velocity data. This technique is sensitive to
  the imbalance of mass at the L4/L5 points of the planet-star
  orbit. No companions were detected above 2$\sigma$ confidence. 
  The median 2$\sigma$ upper limit is 56~$M_\earth$,
  and the most constraining limit is  2.8 $M_\earth$ for the case of
  GJ~436. A similar survey using forthcoming data from the {\it
    Kepler}\, satellite mission, along with the radial-velocity data
  that will be needed to confirm transit candidates, will be sensitive
  to 10--50~$M_\earth$ Trojan companions in the habitable zones of
  their parent stars. As a by-product of this study, we present
  empirical constraints on the eccentricities of the planetary orbits,
  including those which have previously been assumed to be
  circular. The limits on eccentricity are of interest for
  investigations of tidal circularization and for bounding possible
  systematic errors in the measured planetary radii and the predicted
  times of secondary eclipses.

\end{abstract}


\section{Introduction}

Trojan companions are bodies in a 1:1 mean-motion resonance with a planet, 
librating around one of the two triangular Lagrange points (L4 and L5) of the planet's orbit 
around the star. The archetypal example is the population of
Trojan asteroids in resonance with Jupiter. Trojan companions to
Neptune and Mars have also been detected (Sheppard and Trujillo 2006,
Rivkin et al.~2007). Another interesting
example is the pair of Saturnian satellites Calypso and Telesto,
which are in 1:1 resonance with their fellow satellite Tethys (Reitsema 1981).
The presence of Trojan companions and their
orbital and physical characteristics have been considered as
clues to processes in planet formation and migration. Several recent
studies have examined the capture and survival of Trojans in the context
of suspected changes in the orbital architecture of the Solar system
(Morbidelli et al.~2005, Chiang and Lithwick 2005, Kortenkamp et
al.~2004).

Although the Trojan-to-planet mass ratios in the Solar system are very
small ($m_T/m_P \sim 10^{-7}$ for Jupiter), it is conceivable that
Trojans with much higher mass ratios exist in exoplanetary systems. For
circular orbits, even very massive Trojans can be dynamically
stable. Laughlin \& Chambers (2002) explored the viability of Trojans
with mass ratios of unity (i.e., co-orbital planets of equal mass),
finding that such configurations can be dynamically stable over time
scales comparable to or longer than stellar lifetimes. More generally,
the stability of the L4/L5 points depends on the orbital eccentricity
and the relative masses of the Trojan, planet, and star (see, e.g.,
Nauenberg 2002, Dvorak et al.~2004). For many of the known exoplanets,
considerations of dynamical stability allow for massive Trojan
companions. For example, at least 7 of the known gas giant planets
that are within the habitable zones of their parent stars could have
dynamically-stable, terrestrial-mass Trojan companions (Schwarz et
al.~2007).

Several methods have been proposed to detect Trojan companions to
exoplanets. A Trojan may be massive enough to perturb the stellar
motion by an amount that is detectable in the radial-velocity (RV)
orbit of the star (Laughlin \& Chambers 2002). A Trojan in a nearly
edge-on orbit may be large enough for its transit to be detected
photometrically (see, e.g., Croll et al.~2007). For a transiting
planet, the gravitational perturbations from a Trojan companion may
cause a detectable pattern in the recorded transit times (Ford \&
Holman 2007). Alternatively, Ford \& Gaudi (2006) proposed comparing
the measured transit times with the times that would be expected based
only on the RV data and the assumption of a two-body orbit.

An important virtue of the latter technique is that a sensitive search
for Trojans can be performed using only the RV and photometric data
that are routinely obtained while confirming transit candidates and
characterizing the planets. This is in contrast to the first three
methods, for which it is generally necessary to gather new and highly
specialized data (very precise RVs, continuous space-borne
photometry, and a long sequence of precisely-measured transit times,
respectively). For example, Ford \& Gaudi~(2006) and Narita et
al.~(2007) placed upper limits on Trojan companions of approximately
Neptune mass to the transiting planets HD~209458b, HD~149026b and
TrES-1b, using data gathered for other purposes.

In this paper, we present a search for Trojan companions to
25 known transiting exoplanetary systems for which suitable data are
available, using the method of Ford \& Gaudi (2006, hereafter,
``FG''). This paper is organized as follows. The method is described
in \S~2. The compilation and analysis of the data is described in
\S~3. The results are given in \S~4. These results are summarized and
discussed in \S~5, which also looks ahead to the prospects for a
similar search using data from the {\it Kepler}\, mission (Borucki et al.~2008).

As will be explained in \S~2, the orbital eccentricity of the
planet-star orbit affects the interpretation of the data. Hence, a
necessary part of our analysis was the determination of the orbital
eccentricity for each system, or
the justification of the common assumption that the orbit is circular
due to tidal effects. These issues are investigated systematically in
\S~3. Our findings may be of interest independently of our results on
Trojan companions, not only because of the connection to the theory of
tidal circularization, but also because the orbital eccentricity
affects estimates of the planetary radius via transit photometry, as
well as the predicted times of planetary occultations (secondary
eclipses). We discuss these points in \S~5.

\section{Method}
\label{sec:method}

The basic idea of the FG method is to compare the measured transit
time with the expected transit time that is calculated by fitting a
two-body Keplerian orbit to the RV data. We will denote by $t_O$
the observed transit time, and by $t_C$ the calculated transit time,
in which the calculation is based on fitting a two-body Keplerian orbit
to the RV data.
The presence of a Trojan companion as a third body
would cause a timing offset $\Delta t = t_O - t_C$.

This is most easily understood for the case of a planet on a circular
orbit. In such a case, if there is no Trojan companion, the force
vector on the star points directly at the planet, and the observed
transit time $t_O$ coincides with the time $t_V$ when the orbital
velocity of the star is in the plane of the sky (i.e., the time
corresponding to the null in the RV variation). If instead there is a
single Trojan located at L4 or L5 (or librating
with a small amplitude), then the force vector on the star does not
point directly at the planet; it is displaced in angle toward the
Trojan companion, given by $\tan(\phi) \simeq \sqrt{3} 
\epsilon/(2-\epsilon)$ where, $\epsilon = m_T/(m_P + m_T)$ for a Trojan mass
$m_T$ and a planet mass $m_P$  (Ford \& Gaudi 2006). As a result, 
$t_O$ occurs earlier or later than $t_V$, and the time difference is 
given by $\Delta t = \pm \phi P/2 \pi$. 
For small values of the Trojan-to-planet mass ratio,  the magnitude of $t_O - t_V$ is 
proportional to $m_T$, (Ford \& Gaudi 2006):
\begin{equation}
\label{eq:deltat-mt-circular}
\Delta t \simeq \pm 37.5~\text{min}~\bigg( \frac{P}{3 \, \textrm{days}} \bigg)
\bigg( \frac{m_T}{10 \, M_\earth} \bigg)
\bigg( \frac{0.5\,M_{\rm Jup}}{m_P + m_T} \bigg) .
\end{equation}
The positive sign corresponds to a mass excess at the L4 point
(leading the planet) while the negative sign corresponds to a mass
excess at the L5 point (lagging the planet). Thus, 
given a $\Delta t$, the mass excess  can be estimated using 
Eq.~(\ref{eq:deltat-mt-circular}), assuming small Trojan-to-planet mass ratio. 


More generally, the mass excess is given by: 
\begin{equation}
\label{eq:deltat-mt-x}
m_T = m_P \bigg(\frac{2\, \tan(2\pi \Delta t/P)}{\sqrt{3} - |\tan(2\pi \Delta t/P)|}\bigg).
\end{equation}

For an eccentric two-body orbit, the transit time does not generally
coincide with the time of null RV variation, and hence in general $t_C
\neq t_V$. To first order, $t_C-t_V \approx (e\,\cos\omega)P/2\pi$,
where $e$ is the eccentricity and $\omega$ is the argument of
pericenter, and hence one may use the statistic $\Delta t = t_O - t_V
- (e\,\cos\omega)P/2\pi$ to search for Trojan companions. This is how
the problem was described by FG, although we find it useful to cast
the problem more generally as a comparison between $t_O$ and $t_C$. We
emphasize here that $t_O$ depends solely on photometric observations
of transits, while $t_C$ depends almost entirely on RV
observations.\footnote{As explained in \S~\ref{sec:data}, the only
  sense in which $t_C$ depends on photometric data is that we used the
  photometrically-determined orbital period $P$ when fitting the RV
  data, to reduce the number of free parameters.}

Specifically, one calculates $t_C$ by fitting a two-body Keplerian
orbit to the RV data and calculating the expected transit time based
on the the fitted orbital parameters (see, e.g., Kane et
al.~2008). The true anomaly ($f$) corresponding to the transit time is
\begin{equation}
f = \frac{\pi}{2} - \omega,
\end{equation}
from which the eccentric anomaly $E$ can be calculated
using
\begin{equation}
\tan \frac{E}{2} = \sqrt{\frac{1-e}{1+e}} \, \tan \frac{f}{2},
\end{equation}
which in turn leads to the mean anomaly $M$ of the transit
using Kepler's equation,
\begin{equation}
M = E - e\sin E.
\end{equation}
Finally, the calculated transit time $t_C$ is obtained from the
definition of the mean anomaly, $M = 2\pi(t - t_P)/P$, where $t_P$ is
the time of pericenter passage.

Our basic procedure is therefore to determine $t_O$ from published
transit ephemerides, calculate $t_C$ by fitting a two-body Keplerian
orbit to the available RV data, and calculate $\Delta t = t_O - t_C$.
For circular orbits, we use Eq.~(\ref{eq:deltat-mt-x}) to 
determine the Trojan companion mass $m_T$ corresponding to a given 
value of $\Delta t$. For eccentric orbits, the relationship between
$t_C$ and $m_T$ is determined using direct numerical integrations of
3-body systems using a Bulirsch-Stoer algorithm (Varadi et
al.~1996). In these integratons, we hold fixed $P$, $e$, $\omega$, and
the stellar mass $m_S$ at the values given in the literature, and
select a Trojan mass $m_T$ and planetary mass $m_P$ such that the RV
semi-amplitude (Nauenberg~2002)
\begin{equation}
K = \bigg(\frac{2\,\pi\,G}{P}\bigg)^{1/3} \frac{\sqrt{m_P^2+m_T^2+m_Pm_T}}{(m_S + m_P + m_T)^{2/3}\sqrt{1-e^2}},
\label{eqn:3body}
\end{equation}
is equal to the observed value. Hence we simulate the case in which
the observed RV variation is due to the combined force of a planet and
a Trojan, rather than a planet alone, but the RV data alone are
insufficiently precise to discern the difference.\footnote{We verified
  that this discernment is indeed impossible, for $m_T/m_P < 0.5$, for
  the systems with eccentric orbits considered in this paper.} We
compute the transit time $t_C$, repeat the analysis for an increasing
sequence of $m_T$, and fit a polynomial function to the resulting
relationship $t_C(m_T)-t_C(0)$. We found a quadratic
  function, $m_T = a_1 \Delta t + a_2 (\Delta t)^2$, to give a good
  fit to the results.  Taking $m_T$ to be in Earth masses and $\Delta
  t$ in minutes, the coefficients ($a_1$,$a_2$) are
  (0.044,-1.17$\times$ 10$^{-5}$) for GJ 436b and (6.787,-0.001) for
  XO-3b. For the cases of HAT-P-2b and HD 17156b, we find that even
  very low-mass Trojan companions are dynamically unstable, owing to
  the large orbital eccentricities (see \S~\ref{subsec:dynamical}).
  Thus, for those systems, the requirement of dynamical stability is
  more constraining than the empirical upper limit on $m_T$ based on
  the FG method. (As will be described in \S~4, this also proved to be
  true for XO-3b based on the current data.)

\begin{deluxetable*}{l c c c r c c}[htb]
\tablewidth{\textwidth}
\tabletypesize{\scriptsize}
\tablecaption{Description of Data}
\label{tab:rvdata}
\tablehead{\colhead{System} & \colhead{$N_v$\tablenotemark{a}} & \colhead{Jitter\,[m s$^{-1}$]} & \colhead{$\sigma_v$\,[m s$^{-1}$]}
& \colhead{K\,[m s$^{-1}$]} &  \colhead{$\xi=1/\sigma_{(m_T/m_P)}$} & \colhead{References}}
\startdata
HD~209458&	55	&	1.4	&	4.9	&	83.3	&	77.7	&	1,2	\\
HD~17156&	24,8	&	3.3,3.4	&	3.7,6.2	&	273.4	&	72.0	&	47,48,49,50\\
HAT-P-7	&	8	&	6.5	&	6.7	&	213.6	&	55.3	&	51	\\
HD~189733&	16,44	&	0.0,0.0	&	12.0,3.0&	201.3	&	40.6	&	4	\\
TrES-2	&	11,5	&	0.0,0.0	&	7.5,5.8	&	181.5	&	32.2	&	18,19	\\
HAT-P-3	&	9	&	5.1	&	5.8	&	98.7	&	31.3	&	10	\\
HAT-P-4	&	9	&	4.4	&	5.0	&	80.8	&	29.6	&	11	\\
WASP-3	&	6	&	0.0	&	13.8	&	247.7	&	26.9	&	27	\\
WASP-5	&	11	&	11.6	&	21.8	&	276.4	&	25.8	&	29	\\
TrES-3	&	11	&	30.0	&	31.6	&	370.4	&	23.8	&	20	\\
HAT-P-6	&	13	&	10.1	&	11.2	&	116.2	&	23.0	&	13	\\
GJ436	&	52	&	3.4	&	4.2	&	18.3	&	19.4	&	45,46	\\
WASP-4	&	13	&	22.3	&	28.0	&	240.3	&	18.9	&	28	\\
HD~149026&	16	&	5.4	&	6.1	&	46.4	&	18.7	&	3	\\
HAT-P-2	&	13,10,7	&	34.8,88.7,21.4&	35.5,104.5,24.7&980.0	&	16.8	&	7,8,9	\\
WASP-2	&	7	&	16.6	&	17.7	&	157.6	&	14.4	&	25,26	\\
WASP-1	&	7,5	&	2.7,0.0	&	3.9,13.3&	127.8	&	12.8	&	22,23,24\\
XO-3	&	10,10	&	0.0,0.0	&	171.3,159.0&	1486.2	&	12.3	&	34	\\
HAT-P-1	&	15,8	&	4.4,0.0	&	6.4,7.4	&	59.0	&	11.9	&	5,6	\\
TrES-4	&	4	&	0.0	&	10.8	&	98.3	&	11.1	&	21	\\
TrES-1	&	7,8,5	&	0.0,0.0,0.0&	12.1,14.6,3.4&	112.0	&	9.7	&	14,15,16,17\\
CoRoT-Exo-1&	9	&	34.0	&	47.5	&	190.9	&	7.4	&	52	\\
CoRoT-Exo-2&	8,4,3,9	&	60.0,60.0,0.0,0.0&68.4,61.7,27.7,19.0&594.4&	7.3	&	53,54	\\
XO-2	&	9	&	15.6	&	25.0	&	84.1	&	6.2	&	33	\\
HAT-P-5	&	8	&	33.4	&	37.7	&	134.0	&	6.2	&	12	\\
OGLE-TR-182&	20	&	29.7	&	59.5	&	120.0	&	5.5	&	43	\\
OGLE-TR-113&	8	&	83.5	&	93.1	&	286.1	&	5.3	&	41	\\
OGLE-TR-211&	20	&	23.7	&	55.4	&	82.0	&	4.1	&	44	\\
OGLE-TR-56&	11	&	89.0	&	153.1	&	268.3	&	3.6	&	37,38	\\
OGLE-TR-111&	8	&	0.0	&	40.2	&	78.0	&	3.4	&	39,40	\\
OGLE-TR-132&	5	&	51.0	&	68.7	&	167.0	&	3.3	&	42	\\
OGLE-TR-10&	9	&	0.0	&	63.2	&	80.0	&	2.3	&	35,36	\\
XO-1	&	4,6	&	0.0,0.0	&	65.1,16.8&	120.1	&	2.2	&	30,31,32\\
\enddata

\tablerefs{(1)~Laughlin et al.~2005a;\,(2)~Winn et al.~2005;\,(3)~Wolf et al.~2007;\,(4)~Winn et al.~2007a;\,(5)~Bakos et al.~2007a;\,(6)~Winn et al.~2007b;\,(7)~Bakos et al.~2007b;\,(8)~Winn et al.~2007c;\,(9)~Loeillet et al.~2007;\,(10)~Torres et al.~2007a;\,(11)~Kovacs et al.~2007;\,(12)~Bakos et al.~2007c;\,(13)~Noyes et al.~2008;\,(14)~Alonso et al.~2004;\,(15)~Laughlin et al.~2005b;\,(16)~Narita et al.~2007a;\,(17)~Winn et al.~2007d;\,(18)~O'Donovan et al.~2007a;\,(19)~Holman et al.~2007;\,(20)~O'Donovan et al.~2007b;\,(21)~Mandushev et al.~2007;\,(22)~Collier Cameron et al.~2007;\,(23)~Stempels et al.~2007;\,(24)~Charbonneau et al.~2007;\,(25)~Winn et al.~2008;\,(26)~Charbonneau et al.~2007;\,(27)~Pollaco et al.~2007;\,(28)~Wilson et al.~2008;\,(29)~Anderson et al.~2008;\,(30)~McCullough et al.~2006;\,(31)~Holman et al.~2006;\,(32)~Wilson et al.~2006;\,(33)~Burke et al.~2007;\,(34)~Johns-Krull et al.~2007;\,(35)~Konacki et al.~2005;\,(36)~Pont et al.~2007a;\,(37)~Torres et al.~2004;\,(38)~Pont et al.~2007b;\,(39)~Pont et al.~2007c;\,(40)~Winn et al.~2007e;\,(41)~Bouchy et al.~2004;\,(42)~Bouchy et al.~2004;\,(43)~Pont et al.~2007d;\,(44)~Udalski et al.~2008;\,(45)~Maness et al.~2007;\,(46)~Gillon et al.~2007;\,(47)~Fisher et al.~2007;\,(48)~Narita et al.~2007b;\,(49)~Gillon et al.~2007;\,(50)~Irwin et al.~2008;\,(51)~Pal et al.~2008;\,(52)~Barge et al.~2008;\,(53)~Alonso et al.~2008;\,(54)~Bouchy et al.~2008.\,}

\tablenotetext{a}{Multiple values represent multiple data sets available for the system.} 

\end{deluxetable*}


\section{Data Analysis}
\label{sec:data}

The RV data were taken from the available literature on each
system. The references are given in Table~1. These data were generally
obtained for the purpose of discovering or confirming the planet,
although in a few cases the data were obtained for other reasons, such
as precisely measuring the orbital eccentricity (Laughlin et al.~2005)
or for measuring the Rossiter-McLaughlin effect (Winn et al.~2006).
Regarding the latter, the data that were obtained while a transit was
in progress were not used, to avoid the needless complication of
incorporating the Rossiter-McLaughlin effect into the RV model.
However, in those cases the investigators usually gathered additional
data outside of the transit which are useful for refining the
spectroscopic orbit.

Our RV model for an eccentric Keplerian orbit has $4
  + N$ free parameters, where $N$ is the number of independent data
  sets.  Those parameters are the projected planet mass ($m_P\sin i$),
  orbital eccentricity ($e$), argument of pericenter ($\omega$),
  calculated time of midtransit ($t_C$), and a constant additive
  velocity ($\gamma$) for each data set. In practice we use
parameters $e\cos\omega$ and $e\sin\omega$ instead of $e$ and $\omega$
because for small $e$, the errors in $e\cos\omega$ and $e\sin\omega$
are uncorrelated (see, e.g., Winn et al. 2005, Shen \& Turner 2008).
The orbital period $P$ is held fixed at the photometrically determined
value, but of course the transit time $t_C$ is not constrained by the
photometric data, since it is the difference between $t_C$ and the
actual transit time $t_O$ that we are trying to measure. In two cases
for which a long-term acceleration has been identified in the RV data
(GJ 436b and CoRoT-Exo-1b), we include an additional free parameter,
$\dot{\gamma}$\footnote{For the remaining systems, we
    have assumed that the acceleration term is zero. Any real
    acceleration (and any other failures of the single-Keplerian
    model) will appear as ``noise'' in our analysis and will be
    reflected in a larger estimate of stellar ``jitter'' (see
    \S~\ref{subsec:jitter}). We find that our results for $\Delta t$
    do not depend much on whether or not a long-term acceleration  is 
     allowed as an additional parameter, because the error in this parameter is not
    strongly correlated with the error in $\Delta t$.} We assign a
different $\gamma$ to each RV data set, to allow for
telescope-specific velocity offsets. The stellar masses for the
systems were taken from the homogeneous analysis of Torres et
al.~(2008) when possible, and otherwise from the discovery paper.

\subsection{Estimation of Jitter}
\label{subsec:jitter}

For each system, we first analyzed the data with the goal of
reproducing the quoted results in the literature. We fitted a
Keplerian model to the RV data by minimizing the $\chi^2$ statistic
using the AMOEBA algorithm (Press et al.~1992).  The initial
conditions for the free parameters were taken to be the literature
values, and for consistency, for this step we used the exact same choices 
of $P$ and $m_S$ as in the literature. We define $\chi^2$ as
\begin{equation}
\chi^2 = \sum_{i = 1}^{N_{v}} \bigg(   \frac{v_{i,O} - v_{i,C}}{\sigma_i}  \bigg)^2,
\label{eq:chisqr}
\end{equation}
where $v_{i,O}$ and $v_{i,C}$ are the observed and calculated radial
velocities, respectively, and $\sigma_i$ is the corresponding
uncertainty. The uncertainty should include the statistical
uncertainty $\sigma_{\rm stat}$, as well as the systematic error
$\sigma_{\rm sys}$ due to unmodeled instrumental systematic errors and
intrinsic variations of the stellar photosphere, often referred to as
``stellar jitter'' in this context (Wright 2006). To estimate the
appropriate values of $\sigma_i$ for this project, we determined the
value of $\sigma_{\rm sys}$ such that $\chi^2/N_{\rm dof} = 1$ when
using
\begin{equation}
\sigma_i = \sqrt{\sigma_{\rm stat}^2+\sigma_{\rm sys}^2}
\end{equation}
in Eq.~(\ref{eq:chisqr}). Our estimates of the stellar jitter using
this procedure are given in Table~1.

\begin{deluxetable*}{l c c c c c c}
\tabletypesize{\scriptsize}
\tablewidth{\textwidth}
\tablecaption{Inferred orbital eccentricities and related parameters}
\label{tab:ecc}
\tablehead{ \colhead{System}&\colhead{$e\cos\omega$}&\colhead{$e\sin\omega$}&\colhead{$e$\tablenotemark{a}}&\colhead{$\tau_\star$\tablenotemark{b}}&\colhead{$\tau_{\rm circ}$\tablenotemark{c}}&\colhead{$\tau_\star/\tau_{\rm circ}$}}
\startdata
    CoRoT-Exo-1 & $ +0.011_{ -0.071}^{ +0.038}$ & $ -0.073_{ -0.135}^{ +0.133}$ & $<  0.284$ & $   8.00$ & $   0.00$ & $3179.24$  \\
    CoRoT-Exo-2 & $ -0.009_{ -0.025}^{ +0.020}$ & $ +0.054_{ -0.027}^{ +0.025}$ & $<  0.101$ & $   0.50$ & $   0.02$ & $  29.86$  \\
         GJ~436 & $ +0.134_{ -0.006}^{ +0.006}$ & $ -0.016_{ -0.045}^{ +0.045}$ & $  0.138_{ -0.007}^{ +0.013}$ & $   6.00$ & $   1.15$ & $   5.23$  \\
        HAT-P-1 & $ +0.009_{ -0.029}^{ +0.021}$ & $ +0.008_{ -0.049}^{ +0.048}$ & $<  0.099$ & $   2.70$ & $   0.37$ & $   7.30$  \\
        HAT-P-2 & $ -0.516_{ -0.006}^{ +0.005}$ & $ -0.059_{ -0.016}^{ +0.014}$ & $  0.520_{ -0.005}^{ +0.004}$ & $   2.60$ & $  57.59$ & $   0.05$  \\
        HAT-P-3 & $ +0.023_{ -0.053}^{ +0.053}$ & $ +0.033_{ -0.103}^{ +0.062}$ & $<  0.194$ & $   1.50$ & $   0.31$ & $   4.78$  \\
        HAT-P-4 & $ -0.013_{ -0.014}^{ +0.026}$ & $ -0.054_{ -0.040}^{ +0.054}$ & $<  0.123$ & $   4.60$ & $   0.09$ & $  50.18$  \\
        HAT-P-5 & $ +0.026_{ -0.095}^{ +0.095}$ & $ -0.039_{ -0.228}^{ +0.105}$ & $<  0.442$ & $   2.60$ & $   0.09$ & $  27.98$  \\
        HAT-P-6 & $ +0.003_{ -0.023}^{ +0.016}$ & $ +0.042_{ -0.034}^{ +0.034}$ & $<  0.101$ & $   2.30$ & $   0.32$ & $   7.26$  \\
        HAT-P-7 & $ -0.006_{ -0.013}^{ +0.012}$ & $ +0.000_{ -0.019}^{ +0.016}$ & $<  0.038$ & $   2.20$ & $   0.05$ & $  47.88$  \\
      HD~149026 & $ -0.001_{ -0.001}^{ +0.001}$ & $ +0.109_{ -0.068}^{ +0.042}$ & $<  0.179$ & $   1.90$ & $   1.10$ & $   1.73$  \\
       HD~17156 & $ -0.348_{ -0.011}^{ +0.009}$ & $ +0.573_{ -0.006}^{ +0.006}$ & $  0.669_{ -0.007}^{ +0.008}$ & $   5.70$ & $2152.22$ & $   0.00$  \\
      HD~189733 & $ +0.001_{ -0.000}^{ +0.000}$ & $ -0.005_{ -0.011}^{ +0.012}$ & $<  0.024$ & $   6.80$ & $   0.05$ & $ 136.62$  \\
      HD~209458 & $ +0.001_{ -0.002}^{ +0.002}$ & $ +0.008_{ -0.014}^{ +0.011}$ & $<  0.028$ & $   3.10$ & $   0.11$ & $  27.22$  \\
     OGLE-TR-10 & $ +0.245_{ -0.886}^{ +0.429}$ & $ +0.436_{ -0.264}^{ +0.357}$ & $<  1.000$ & $   3.20$ & $   0.13$ & $  25.54$  \\
    OGLE-TR-111 & $ +0.163_{ -0.028}^{ +0.616}$ & $ +0.099_{ -0.559}^{ +0.072}$ & $<  0.964$ & $   8.80$ & $   0.51$ & $  17.39$  \\
    OGLE-TR-113 & $ -0.044_{ -0.087}^{ +0.092}$ & $ +0.152_{ -0.187}^{ +0.104}$ & $<  0.417$ & $  13.20$ & $   0.01$ & $1204.61$  \\
    OGLE-TR-132 & $ +0.247_{ -0.198}^{ +0.529}$ & $ +0.279_{ -0.477}^{ +0.084}$ & $<  0.993$ & $   1.20$ & $   0.02$ & $  72.76$  \\
    OGLE-TR-182 & $ -0.071_{ -0.448}^{ +0.064}$ & $ +0.352_{ -0.213}^{ +0.147}$ & $<  0.960$ & $   2.00$ & $   0.83$ & $   2.41$  \\
    OGLE-TR-211 & $ +0.007_{ -0.170}^{ +0.130}$ & $ +0.144_{ -0.244}^{ +0.244}$ & $<  0.858$ & $   2.00$ & $   0.15$ & $  12.91$  \\
     OGLE-TR-56 & $ +0.003_{ -0.790}^{ +0.279}$ & $ +0.519_{ -0.399}^{ +0.251}$ & $<  0.998$ & $   3.20$ & $   0.00$ & $1139.56$  \\
         TRES-1 & $ +0.003_{ -0.002}^{ +0.002}$ & $ -0.039_{ -0.028}^{ +0.030}$ & $<  0.084$ & $   3.70$ & $   0.19$ & $  19.75$  \\
         TRES-2 & $ +0.022_{ -0.017}^{ +0.015}$ & $ -0.024_{ -0.027}^{ +0.019}$ & $<  0.078$ & $   5.00$ & $   0.07$ & $  75.20$  \\
         TRES-3 & $ +0.028_{ -0.018}^{ +0.016}$ & $ -0.031_{ -0.040}^{ +0.036}$ & $<  0.101$ & $   0.60$ & $   0.00$ & $ 137.03$  \\
         TRES-4 & $ +0.199_{ -0.642}^{ -0.056}$ & $ +0.434_{ -0.590}^{ -0.054}$ & $<  0.859$ & $   2.90$ & $   0.05$ & $  54.60$  \\
         WASP-1 & $ +0.006_{ -0.038}^{ +0.031}$ & $ +0.009_{ -0.039}^{ +0.035}$ & $<  0.088$ & $   3.00$ & $   0.02$ & $ 129.19$  \\
         WASP-2 & $ -0.231_{ +0.007}^{ +0.331}$ & $ -0.016_{ -0.387}^{ +0.058}$ & $<  0.547$ & $   5.60$ & $   0.05$ & $ 116.01$  \\
         WASP-3 & $ -0.012_{ -0.019}^{ +0.027}$ & $ +0.015_{ -0.039}^{ +0.047}$ & $<  0.098$ & $   2.10$ & $   0.02$ & $  93.68$  \\
         WASP-4 & $ +0.009_{ -0.024}^{ +0.021}$ & $ +0.018_{ -0.042}^{ +0.047}$ & $<  0.096$ & $   2.00$ & $   0.00$ & $ 931.94$  \\
         WASP-5 & $ +0.032_{ -0.017}^{ +0.017}$ & $ +0.026_{ -0.033}^{ +0.027}$ & $<  0.088$ & $   2.00$ & $   0.03$ & $  77.79$  \\
           XO-1 & $ +0.017_{ -0.149}^{ +0.064}$ & $ +0.066_{ -0.191}^{ +0.099}$ & $<  0.290$ & $   1.00$ & $   0.43$ & $   2.32$  \\
           XO-2 & $ +0.014_{ -0.056}^{ +0.056}$ & $ -0.216_{ -0.177}^{ +0.160}$ & $<  0.515$ & $   5.80$ & $   0.12$ & $  48.43$  \\
           XO-3 & $ +0.217_{ -0.016}^{ +0.016}$ & $ -0.063_{ -0.031}^{ +0.034}$ & $  0.229_{ -0.018}^{ +0.016}$ & $   2.70$ & $   2.98$ & $   0.91$  \\
\enddata

\tablenotetext{a}{ 95.4 \% confidence limits on eccentricity. For four systems which are clearly eccentric, the mode and the 68.3\% 
confidence  limits are reported.} 

\tablenotetext{b}{Nominal age of the system in Gyr\,: Taken from Torres et al. 2008, when possible, and from
the discovery papers for systems not analyzed by Torres et al. 2008. For Corot-Exo-1, we assume a nominal age of 8 Gyr\, based on the 
reasoning that is has to be a fairly old main-sequence star (see Barge et al. 2008).} 

\tablenotetext{c}{Circularization time-scale (in Gyr) for the planetary orbit, assuming $Q = 10^6$ (see the text).} 

\end{deluxetable*}

\subsection{Data Selection}

Our desire was to perform as wide a search as possible, using all
publicly available data, but for some systems the RV data is so sparse
that meaningful constraints cannot yet be obtained. To guide our
selection of systems, we used a figure-of-merit based on the results
of the Fisher information analysis presented by FG.  Those authors
showed that in the limit of continuous RV sampling with uniform
errors, the uncertainty in $\Delta t$ will approach $\sigma_{\Delta t}
= (1/2 \pi^2 N_v)^{1/2}P\,\sigma_v/K$, where $N_v$ is the number of
radial-velocity data points and $\sigma_v$ is the error per point.
For a circular orbit the corresponding uncertainty in $m_T$ will
approach $\sigma_{m_T} = (8/3 N_v)^{1/2} m_P\,\sigma_{v}/K$. Thus, it
is possible to estimate the expected uncertainty in $\Delta t$ and
$m_T$, given the published system parameters, without fitting the
actual data. We calculated the figure of merit
\begin{equation}
\xi \equiv \frac{\sqrt{(3N_v/8)}\,K}{\sigma_v} \approx 1/\sigma_{(m_T/m_P)}
\label{eq:goodness}
\end{equation}
for each system in the literature at the outset of this project, and ranked the 
systems accordingly. Table~1 shows $\xi$ for all the systems, along with 
$N_v$, $\sigma_v$, and $K$. For systems where multiple data sets are available, the effective
$\sigma_{v}/N_{v}^{1/2}$ is obtained by adding in quadrature the
corresponding terms from the different data sets. A higher value of
$\xi$ reflects better quality of data. From MCMC analyses of all the
systems, we find that for systems with $\xi < 6$, the
fitting algorithm is susceptible to poor convergence and allows for
unphysical parameter ranges. Since the scientific return on such
low-$\xi$ systems is comparatively poor we decided to remove them from
consideration rather than eliminate these fitting problems.
In what follows we focus exclusively on the 25 systems with $\xi > 6$.

\subsection{Assumptions for orbital eccentricity}
\label{subsec:ecc}

As explained in \S~2, the orbital eccentricity affects the calculation
of $t_C$ and also affects the relation between 
$\Delta t$ and the possible Trojan companion mass $m_T$. Hence it is
imperative to consider the possiblity of eccentric orbits. Most of the
currently known transiting planets are in very close orbits, where the
effects of tidal interactions between the star and planet---and
orbital circularization in particular---are expected to be significant
(Rasio et al.~1996, Trilling 2000, Dobbs-Dixon et al.~2004). A common
practice is to assume that, in the absence of positive
evidence for an eccentric orbit, the orbital eccentricity has been
reduced to insignificance by the action of tides.

If the assumption of a circular orbit could be justified, it would be
advantageous for the present study because it would remove 2 free
parameters from the Keplerian model ($e$ and $\omega$) and thereby
strengthen the determination of the other parameters, including the
key parameter $t_C$. Our approach was to assume the orbit to be
circular only when (1) a circular orbit is consistent with the RV
data, (2) the estimated stellar age is more than 20 times larger than
the estimated timescale for tidal circularization, and (3) no
constraint on $e\cos\omega$ is available because no planetary
occultations (secondary eclipses) have been observed. These points are
explained in detail in the paragraphs to follow.

To test whether the RV data are consistent with a circular orbit, we
fitted a Keplerian model to the RV data using a Markov Chain Monte
Carlo (MCMC) technique, employing a Metropolis-Hastings algorithm
within the Gibbs sampler (see, e.g., Tegmark et al.~2004; Ford~2005;
Holman et al.~2006; Winn et al.~2007a). 
For this step, the free parameters were $m_P\sin i$, 
$e\cos\omega$, $e\sin\omega$ and a $\gamma$ for each data set. 
Uniform priors were used for all 
parameters. The fitting statistic, $\chi^2$, was defined in Eq.~(\ref{eq:chisqr}). 
A single chain of $\sim 10^6$ links was used for each system. The jump 
sizes for the various parameters were set such that the acceptance 
rate for each parameter was $\sim$20\%. For each parameter, we found
the mode of the {\it a posteriori}\, distribution (marginalized over
all other parameters), and the 68.3\% confidence interval, defined as
the range that excludes 15.9\% of the probability at each extreme of 
the {\it a posteriori}\, distribution. For cases when
$e\cos\omega$ and $e\sin\omega$ were both consistent with zero, we
also found the 95.4\%-confidence upper limit on $e$. Table~2 gives the
results. All of the systems were found to be consistent with a
circular orbit, except for the 4 well-known eccentric systems GJ~436,
HAT-P-2, HD~17156, and XO-3.

For the estimated timescale for tidal circularization, we used
(Goldreich \& Soter 1966):
\begin{equation}
 \tau_{\rm circ} = \frac{4}{63} Q\bigg(\frac{a^3}{G\,m_S}\bigg)^{1/2} \frac{m_P}{m_S}\bigg(\frac{a}{R_p}\bigg)^5,  
 \label{eq:tcirc}
\end{equation}
which is based on the highly simplified, widely-used model of tidal
dissipation in which the tidal bulge experiences a constant phase lag
due to tidal friction. Here, $a$ is the orbital separation, and $R_p$ is the 
radius of the planet. The dimensionless number $Q$ is inversely
proportional to the dissipation rate.  In the solar system, Jupiter is
thought to have $Q\sim 10^5$ (Ioannou \& Lindzen 1993) to the extent
that this simplified model is applicable. For our purpose, a 
necessary condition for assuming the orbit to be circular was that 
$\tau_\star/\tau_{\rm circ} > 20$, i.e., there
have been at least 20 $e$-foldings of tidal circularization, according
to this model.  In calculating $\tau_{\rm circ}$ we
assumed $Q=10^6$, which is conservative in the sense that a larger $Q$
corresponds to a longer calculated timescale for circularization, and
a smaller risk that we are assuming a circular orbit when this
assumption is not justified.

\begin{figure}[h]
  \begin{center}
 \includegraphics[width=0.5\textwidth]{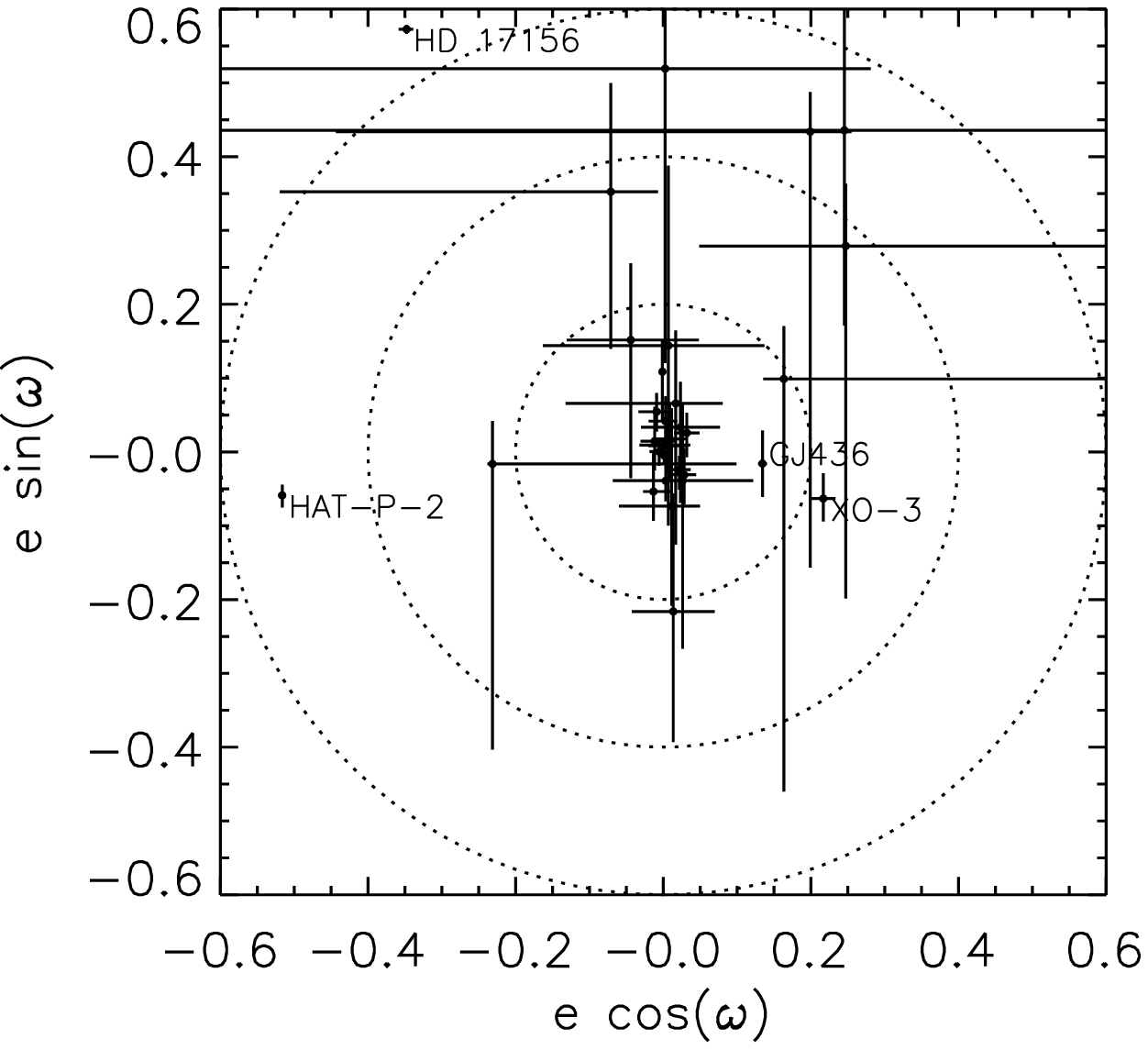}
    \caption{Constraints on $e\cos\omega$ and $e\sin\omega$ based on
      fitting a Keplerian orbit to the RV data, and using constraints
      from the observed time of the secondary eclipse when
      available. The four clearly eccentric systems are labeled. The
      dotted circles show the eccentricity contours corresponding to
      $e=0.2$, 0.4 and 0.6.}
    \label{fig:eccen}
  \end{center}
\end{figure}

\begin{figure}[h]
  \begin{center}
 \includegraphics[width=0.5\textwidth]{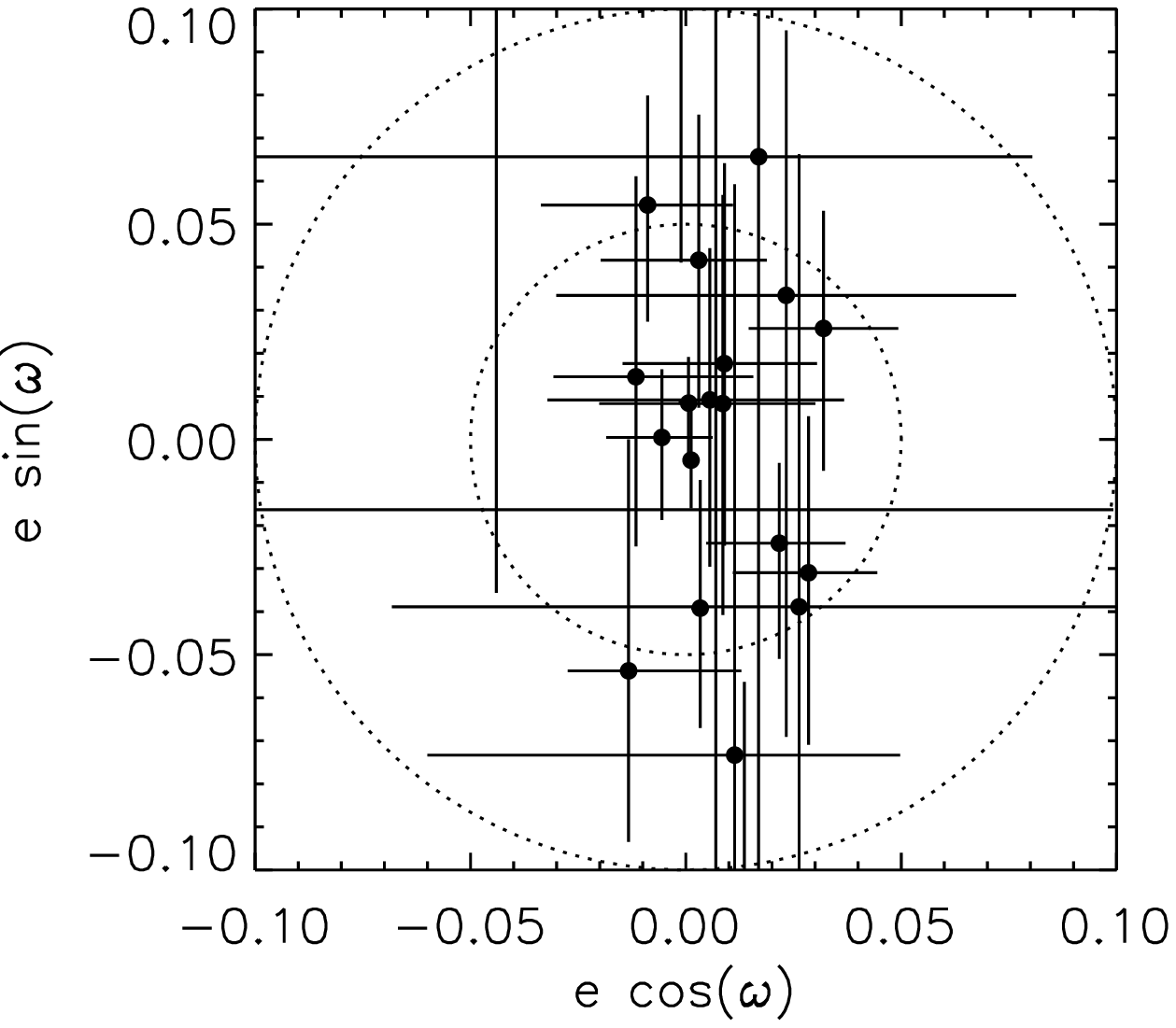}
    \caption{The inner region of Fig~\ref{fig:eccen}. The dotted circles 
     show the eccentricity contours corresponding to $e=0.05$ and 0.1.}
    \label{fig:eccen_zoom}
  \end{center}
\end{figure}

For a few systems, additional constraints on the orbital eccentricity
are available because a planetary occultation (secondary eclipse) has
been observed. For small eccentricities, the variables $e\cos\omega$
and $e\sin\omega$ are directly related to the interval between the
transit and occultation, and the relative durations of those two
events (Kallrath \& Milone 1999, Charbonneau et al.~2005):
\begin{equation}
e\,cos\omega = \frac{\pi}{2P} \bigg( t_{\rm occ} - t_{\rm tra} - \frac{P}{2} \bigg), 
\label{eq:ecosw}
\end{equation}
\begin{equation}
e\,sin\omega = \frac{\Theta_{\rm tra} - \Theta_{\rm occ}}{\Theta_{\rm tra} + \Theta_{\rm occ}}
\end{equation}
where $t_{\rm tra}$ and $t_{\rm occ}$ are the times of transit and
occultation, and $\Theta_{\rm tra}$ and $\Theta_{\rm occ}$ are the
corresponding durations. In all cases to date, the bounds on
$e\sin\omega$ that follow from this relation are weaker than bounds 
from the RV data (see, e.g., Winn et al.~2005), and the 
bounds on $e\cos\omega$ are more constraining.  Thus, for those
cases in which an occultation has been observed, we add a term to our
$\chi^2$ statistic to enforce the corresponding constraint on
$e\cos\omega$:
\begin{equation}
\label{eq:chi2-ecosw}
\chi^2 = \sum_{n = 1}^{N_v} \bigg(   \frac{v_O - v_C}{\sigma_v}  \bigg)^2 +  \bigg(   \frac{(e\,cos\omega)_O - 
(e\,cos\omega)_C}{\sigma_{e\,cos\omega}}  \bigg)^2,
\end{equation}
where $(e\cos\omega)_o$ and $(e\cos\omega)_c$ are the ``observed''
and calculated values of $e\cos\omega$. By ``observed'' we mean the
value that follows from the measured interval between transits and
occultations when inserted into Eq.~(\ref{eq:ecosw}). The 5 systems in
this category, and the constraints on $e\cos\omega$, are listed in
Table~3. In those cases, because such a powerful empirical constraint
is available, we do not assume the orbit to be circular even if the RV
data are consistent with a circular orbit and $\tau_\star/\tau_{\rm circ} >
20$.
 
\begin{deluxetable}{l r r}
\tabletypesize{\small}
\tablecaption{Constraints on $e\,cos\omega$ from Secondary Eclipse Observations}
\label{tab:seclipse}
\tablehead{\colhead{System}&\colhead{$e\,cos\omega$\tablenotemark{a}}&\colhead{Reference\tablenotemark{b}}}
\startdata

HD~209458 &   $0.0002 \pm 0.0021$ &  Deming et al.~2005 \\
HD~189733 &   $0.0012 \pm 0.0002$ &  Knutson et al.~2007 \\
HD~149026 &  $-0.0011 \pm 0.0009$ &  Harrington et al.~2007 \\
TrES-1 &  $0.0030 \pm 0.0019$ &  Charbonneau et al.~2005 \\
GJ~436  &  $0.1346 \pm 0.0059$ &  Deming et al.~2007 \\

\enddata

\tablenotetext{a}{Evaluated from the transit ephemeris and the secondary eclipse time using Eqn. (\ref{eq:ecosw}).}
\tablenotetext{b}{Reference in literature from which the secondary eclipse time was obtained.}

\end{deluxetable}

\subsection{Constraints on the masses of Trojan companions}
\label{subsec:mcmc}

For each system, after determining the appropriate level of stellar
jitter and deciding whether or not the assumption of a circular orbit
is justified, we determined the key parameter $t_C$ and its
uncertainty using the same MCMC code that was described in the
previous section. In all cases the free parameters included $m_P\sin i$,
$\gamma$, and $t_C$. In cases for which the a circular orbit was not
assumed, we also fitted for $e\cos\omega$ and $e\sin\omega$.  For
the special cases of GJ~436 and CoRoT-Exo-1, a velocity gradient
$\dot{\gamma}$ was also included as a free parameter.  The basic
fitting statistic, $\chi^2$, was defined in Eq.~(\ref{eq:chisqr}), and
for the systems in Table~3 an {\it a priori}\, constraint on
$e\cos\omega$ was applied as in Eq.~(\ref{eq:chi2-ecosw}).

To determine the photometric transit time $t_O$, we used the most
precise published photometric ephemeris to compute a predicted transit
time close to the midpoint of the RV time series. We then computed the
key parameter $\Delta t = t_O - t_C$, the difference between the
photometrically observed transit time and the transit time calculated
from the RV data assuming zero Trojan mass. In all cases, the
uncertainty in $t_O$ is negligible in comparison to the uncertainty in
$t_C$. The results for $\Delta t$ are translated into constraints on
the Trojan mass $m_T$ using Eq.~(\ref{eq:deltat-mt-x}) for
circular orbits, and using the numerical integrations described in
\S~\ref{sec:method} for eccentric orbits.

\section{Results}
\label{sec:results}

\begin{deluxetable*}{l c c l c}
\tablewidth{\textwidth}
\tabletypesize{\footnotesize}
\tablecaption{Observational Constraints on Trojan Masses}
\label{tab:trojans}
\tablehead{\colhead{System}&\colhead{$\Delta t $}&\colhead{$m_{T}$}&\multicolumn{2}{c}{Upper bound (95.4 \% confidence)} \\
	\colhead{} & \colhead{[{\rm min}]} & \colhead{$[M_{\earth}]$} &\multicolumn{2}{c}{} \\ 
	\colhead{} & \colhead{} & \colhead{} & \colhead{$m_{T}/M_{\earth}$} & \colhead{$m_{T}/m_{P}$} }
\startdata
    CoRoT-Exo-1 & $+0020.6_{-0042.2}^{+0042.2}$ & $+0018.0_{-0042.1}^{+0061.4}$ & $<0116.5$ & $<0.35$  \\
    CoRoT-Exo-2 & $+0012.8_{-0015.7}^{+0020.5}$ & $+0039.6_{-0048.0}^{+0078.0}$ & $<0153.0$ & $<0.14$  \\
         GJ 436 & $-0005.5_{-0031.0}^{+0034.1}$ & $-0000.4_{-0001.2}^{+0001.8}$ & $<0002.8$ & $<0.12$  \\
        HAT-P-1 & $+0126.2_{-0082.7}^{+0069.6}$ & $+0022.1_{-0014.3}^{+0021.0}$ & $<0052.9$ & $<0.32$  \\
        HAT-P-2 & $-0055.0_{-0031.7}^{+0031.7}$ & $-0146.4_{-0079.0}^{+0086.9}$ & $<0280.5$\tablenotemark{a} & $<0.10$  \\
        HAT-P-3 & $-0139.2_{-0100.6}^{+0138.3}$ & $-0045.7_{-0097.9}^{+0046.4}$ & $<0262.8$ & $<1.25$  \\
        HAT-P-4 & $-0012.5_{-0018.3}^{+0024.4}$ & $-0004.4_{-0007.3}^{+0008.2}$ & $<0016.5$ & $<0.08$  \\
        HAT-P-5 & $-0025.5_{-0108.5}^{+0098.6}$ & $-0011.8_{-0072.1}^{+0057.6}$ & $<0147.7$ & $<0.46$  \\
        HAT-P-6 & $-0037.0_{-0171.0}^{+0141.3}$ & $-0012.6_{-0097.0}^{+0067.9}$ & $<0198.9$ & $<0.59$  \\
        HAT-P-7 & $-0000.6_{-0009.5}^{+0011.2}$ & $-0000.8_{-0012.9}^{+0014.2}$ & $<0026.5$ & $<0.05$  \\
       HD 149026 & $+0026.0_{-0038.1}^{+0028.6}$ & $+0004.7_{-0006.6}^{+0007.5}$ & $<0016.6$ & $<0.14$  \\
       HD 17156 & $-0018.4_{-0084.6}^{+0093.5}$ & $-0004.8_{-0020.3}^{+0024.6}$ & $<0043.8$\tablenotemark{a} & $<0.04$  \\  
      HD 189733 & $-0008.8_{-0010.7}^{+0010.7}$ & $-0007.7_{-0008.8}^{+0009.4}$ & $<0022.1$ & $<0.06$  \\
      HD 209458 & $+0002.3_{-0009.3}^{+0010.8}$ & $+0000.6_{-0002.7}^{+0003.5}$ & $<0006.1$ & $<0.03$  \\
         TrES-1 & $-0004.4_{-0011.2}^{+0012.7}$ & $-0001.6_{-0004.6}^{+0004.6}$ & $<0009.9$ & $<0.04$  \\
         TrES-2 & $-0008.6_{-0012.5}^{+0015.2}$ & $-0006.2_{-0010.6}^{+0010.6}$ & $<0024.8$ & $<0.06$  \\
         TrES-3 & $-0014.4_{-0011.7}^{+0010.6}$ & $-0034.1_{-0028.4}^{+0025.8}$ & $<0081.3$ & $<0.14$  \\
         TrES-4 & $-0125.0_{-0080.9}^{+0089.0}$ & $-0049.3_{-0057.5}^{+0035.4}$ & $<0143.8$ & $<0.49$  \\
         WASP-1 & $-0017.6_{-0049.5}^{+0054.5}$ & $-0007.8_{-0034.4}^{+0031.1}$ & $<0070.5$ & $<0.24$  \\
         WASP-2 & $-0122.9_{-0052.7}^{+0063.8}$ & $-0082.0_{-0070.1}^{+0042.1}$ & $<0199.1$ & $<0.74$  \\
         WASP-3 & $+0016.1_{-0012.9}^{+0011.7}$ & $+0023.8_{-0017.5}^{+0019.3}$ & $<0056.1$ & $<0.10$  \\
         WASP-4 & $-0004.7_{-0013.2}^{+0014.5}$ & $-0006.5_{-0020.0}^{+0020.0}$ & $<0043.0$ & $<0.11$  \\
         WASP-5 & $-0014.1_{-0011.4}^{+0012.6}$ & $-0021.4_{-0019.5}^{+0019.5}$ & $<0054.7$ & $<0.11$  \\
           XO-2 & $+0034.8_{-0091.5}^{+0100.7}$ & $+0009.4_{-0027.4}^{+0047.0}$ & $<0088.2$ & $<0.49$  \\
           XO-3 & $-0057.9_{-0067.2}^{+0068.8}$ & $-0384.4_{-0424.3}^{+0479.6}$ & $<1097.6$\tablenotemark{a} & $<0.26$  \\
\enddata
\tablenotetext{a}{For this system the upper limit due to dynamical stability is more constraining than
that obtained from the data analysis in this paper using the FG method. Dynamical stability constraints do not allow for Trojan companions to HAT-P-2b and HD 17156b. And, for XO-3b, the upper limit due 
to stability is 105 M$_\oplus$.} 

\end{deluxetable*}

\subsection{Constraints on Trojan Masses}

Table~4 gives the 68.3\% (1$\sigma$) confidence intervals for $\Delta
t$ and $m_T$ for all 25 systems under consideration, as well as the
95.4\% (2$\sigma$) upper limits on $m_T$ and $m_T/m_P$. 
In all the cases, the result for $\Delta t$ was consistent 
with zero within 2$\sigma$. The system that was closest to a 2$\sigma$ 
detection was WASP-2, for which $\Delta t = -123^{+64}_{-53}$~minutes. 
The result for WASP-2 is therefore worth following up with additional RV data. 
However, in a sample of 25 systems, even if $\Delta t$ is always 
consistent with zero, one expects approximately one 2$\sigma$ outlier. 
Hence our survey has not produced compelling evidence for a 
Trojan companion in this ensemble.

\begin{figure}[h]
\begin{center}
\includegraphics[width=0.5\textwidth]{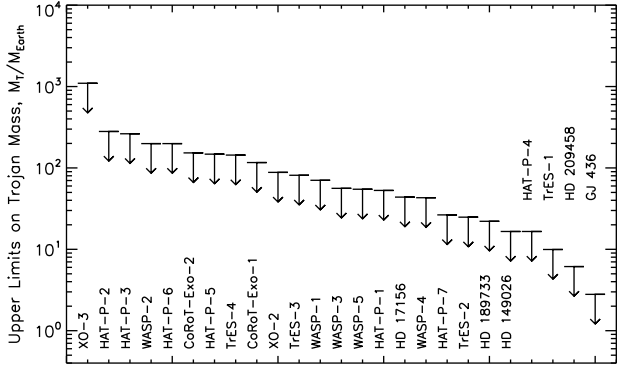}
\caption{95.4\%-confidence upper limits on masses of Trojan
  companions. The systems are ordered from the weakest to the
  strongest upper bound, from left to right.}
\label{fig:trojans}
\end{center}
\end{figure}

\begin{figure}[h]
\begin{center}
\includegraphics[width=0.5\textwidth]{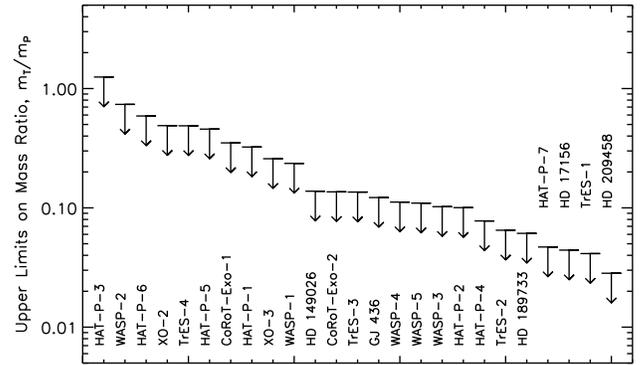}
\caption{95.4\%-confidence upper limits on Trojan-to-planet mass
  ratios. The systems are ordered from the weakest to the strongest
  upper bound, from left to right.}
\label{fig:ratios}
\end{center}
\end{figure}

The 2$\sigma$ upper limits on $m_T$ and on $m_T/m_P$ are shown in
Figures~\ref{fig:trojans} and \ref{fig:ratios}, respectively. The
systems are ordered from least-constrained to best-constrained, going
from left to right. The median upper limit on $m_T$ is 56~$M_\earth$,
with the most constraining limit of  2.8~$M_\earth$ holding for the
Neptune-sized planet GJ~436. Such a powerful upper limit is possible
in this case because of the small stellar and planetary masses, and
the copious RV data that is available for this system. The median
upper limit on the mass ratio $m_T/m_P$ is 0.1.

It is possible to compare our results to those obtained previously for
3 particular systems. For HD~209458, FG found $\Delta t = 13 \pm 9$
minutes and we find $2^{+11}_{-9}$~min. For HD~149026, FG found $\Delta t =
-13\pm 27$~min and we find $26^{+29}_{-38}$~min. For TrES-1, Narita et
al.~(2007) found $\Delta t = - 3.2 \pm 11.8$~min, assuming a circular orbit, 
and we find $-4^{+13}_{-11}$~min, allowing the orbit to be eccentric but using the 
constraint on $e\cos\omega$ from secondary eclipse. These results are all 
consistent with zero with approximately the same range of uncertainty. 
Minor differences in the quoted central
values are probably attributable to minor differences in the fitting
procedures and in reporting median values of the {\it a posteriori}\,
distributions rather than modes. We also find our uncertainties to be
in general agreement with the forecasted uncertainties based on the
Fisher information analysis of FG.

\subsection{Considerations of dynamical stability}
\label{subsec:dynamical}
For a planet on a circular orbit, non-librating
  Trojan companions are stable as long as the masses satisfy the
  condition (Laughlin \& Chambers 2002):
\begin{equation}
\frac{m_P + m_T}{(m_S + m_P + m_T)} \leq 0.03812,
\label{eqn:mt_circ}
\end{equation}
where $m_S$, $m_P$ and $m_T$ are the masses of the star, planet and
Trojan companion, respectively. This criterion allows for Trojan
``companions'' that are just as massive as the planet itself, even for
planets as massive as 10~$M_{\rm Jup}$ around a Sun-like star.

However, the condition for stability of Trojan
  companions depends strongly on the eccentricity of the
  orbit. Nauenberg (2002) reported just such a study, showing the
  stability domain of bodies in 1:1 resonance as a function the
  eccentricity of the orbit and the Routh parameter ($\gamma_R$) given
  by:
\begin{equation}
\gamma_R = \frac{m_S m_P + m_P m_T + m_S m_T }{(m_S + m_P + m_T)^2}.
\label{eqn:gammaR}
\end{equation}
Given the masses of the three bodies, and the eccentricity of the
system, one can calculate $\gamma_R$ and determine from Fig.~5 of
Nauenberg~(2002) whether or not the system is stable in 1:1 resonance.
Conversely, given the eccentricity of the system, Fig.~5 of
Nauenberg~(2002) gives the the maximum $\gamma_R$ allowed for
stability which, along with $m_S$ and $m_P$, gives the maximum Trojan
mass allowed in the system.

  We calculated such limits on the Trojan masses in the
  four eccentric systems that we analyzed in this work. We find the
  mass limits to be zero for HAT-P-2b and HD 17156b (in agreement with
  the results of our 3-body integrations that were described in
  \S~\ref{sec:method}), 105 M$_\oplus$ for XO-3b, and 3030 M$_\oplus$
  for GJ 436b.  Thus, for HAT-P-2b, HD~17156b, and XO-3b, the upper
  limit on $m_T$ based on considerations of dynamical stability is
  more constraining than the empirical upper limit using the FG
  method. For GJ~436b, the upper limit on $m_T$ using the FG method
  ( 2.8 M$_\oplus$) is much stronger than the upper limit imposed by
  the stability requirement.

\section{Discussion}

Exoplanetary science has provided enough surprises that an appropriate
maxim for observers is: If you can look for a novel effect or
phenomenon that is at least physically plausible, then you should do
so, especially when this can be done with existing data. We have
obeyed this maxim by conducting a search for Trojan companions to 25
transiting planets. Specifically we have put the technique of FG into
practice with a much larger ensemble than has been previously
analyzed. We have conducted a search for planets in particular
locations (L4/L5) with a median sensitivity of $\sim$56~$M_\earth$,
without gathering any new data. Instead, we asked the RV data: when
should the transit occur if there is no Trojan companion? Then we
consulted the photometric ephemeris to determine when a transit
actually did occur, and interpreted the time difference as a
measurement or constraint on Trojan companions. 
Our results must be understood as constraints on the imbalance of
mass residing at the L4 and L5 positions. Equally massive Trojan
companions at those positions would produce opposite effects, and no
net FG signal, when averaged over the libration periods.

For some systems such as HAT-P-3 and WASP-2, the existing RV data are
sparse and noisy enough to allow only the barest constraints on Trojan
companions, with masses comparable to the planetary mass. These
constraints are nevertheless physically meaningful, in the sense that
Laughlin \& Chambers (2002) have shown that equal-mass planets in a
1:1 mean-motion resonance and circular orbits can be dynamically stable.
In one case, WASP-2, we found a near - 2$\sigma$ evidence for a
timing offset that could be interpreted as a Trojan companion. This is
not compelling evidence, especially given that we examined a total of
25 systems, but this system is worthy of follow-up. 
In one case, GJ~436, we have found a 2$\sigma$ upper limit of
$ 2.8$~$M_\earth$ on $m_T$. A positive detection at this level would
have represented the least-massive planet detection to date, which is
remarkable considering that we did not gather any new data. In no case
was there evidence for a timing offset at the 2$\sigma$ level.

As explained in \S~\ref{subsec:ecc}, as part of this study we assessed
the justification for assuming that a given planetary orbit is
circular, given the existing RV data and reasonable estimates of the
stellar age and the timescale for tidal circularization. One part of
this assessment was the determination of empirical constraints on
$e\cos\omega$ and $e\sin\omega$ based on the RV data. These results
may be interesting to other investigators, independently of our
results on Trojan companions. The compilation in Table~2 of the
results for $\tau_\star$ and $\tau_{\rm circ}$ may be useful to those
who are interested in making inferences about the tidal
circularization process from the ensemble of transiting planets (see,
e.g., Rasio et al.~1996, Trilling 2000, Dobbs-Dixon et al.~2004,
Jackson et al.~2008, Mazeh 2008). The limits on $e\cos\omega$ are also
useful for bounding the possible error in the predicted times of
occultations (secondary eclipses), using the relationship given in
Eqn.~(\ref{eq:ecosw}). In addition, although the planetary radius that
is determined from transit photometry depends mainly on the observed
transit depth, there is a secondary dependence on the transit
timescales (the total duration, and the duration of ingress or egress)
and the sky-projected orbital speed of the planet during the transit.
The latter quantity is not directly observable; it depends on the
orbital period, the stellar mass, and the orbital eccentricity and
argument of pericenter. Thus there is a secondary dependence of the
inferred planetary radius on $e$ and $\omega$ (see, e.g., Barnes~2007,
or McCullough et al.~2008 for a particular example). The results of
Table~2 can be used to bound the systematic error that could arise
from this effect.

Having completed this survey using the existing transit data, one may
wonder about the achievable limits on Trojan companions using data
from ambitious future transit surveys. We consider here the particular
case of the {\it Kepler}\, satellite mission (Borucki et al.~2008),
whose primary goal is the detection of Earth-like planets in the
habitable zones of Sun-like stars.  {\it Kepler}\, is also likely to
find larger planets such as gas giants in the habitable zones of their
parent stars, and such planets may have lower-mass Trojan companions
that are perhaps ``more habitable'' than the gas giants.
 Such companions might be detectable photometrically
  if they are very nearly coplanar with the transiting planet.
  However, even if they do not transit, they can be detected with the
  FG method. It is therefore natural to ask what constraints on
Trojan companions will be possible for a given planet that {\it
  Kepler}\, detects in the habitable zone of a Sun-like star, using
only the photometric data and the RV data that are routinely gathered
for the purposes of confirming and characterizing transiting planets.

\begin{figure}[h]
\begin{center}
\includegraphics[width=0.5\textwidth]{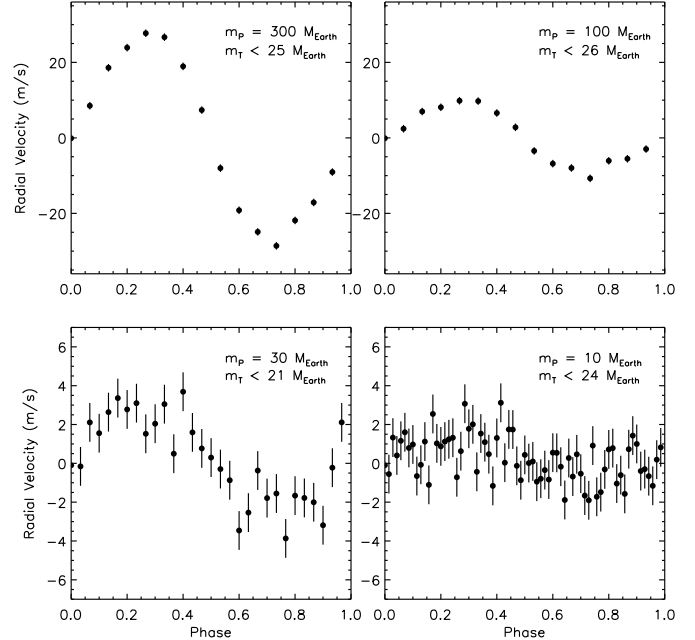}
\caption{Simulated radial velocities and the corresponding 2$\sigma$ upper limits on Trojan companions to potential \textit{Kepler} detections in the habitable zone around a Sun-like star with V = 12.}
\label{fig:kep}
\end{center}
\end{figure}

To answer this question, we consider four different cases, in which
{\it Kepler}\, finds a planet of mass 300~$M_\earth$ (case A),
100~$M_\oplus$ (B), 30~$M_\oplus$ (C) or 10 $M_\oplus$ (D), orbiting a
star of solar mass and apparent magnitude $V=12$, with a period of
1~yr and an orbital eccentricity of 0.1. We simulate RV data for each
system with $\sigma_v = 1$~m~s$^{-1}$, and a number of data points
$N_v$ that seems realistic for the {\it Kepler}\, follow-up
program. For cases A and B we assume $N_v=15$. For case C we
assume $N_v=30$, which is sufficient to measure the planetary mass to
at least 10\% according to the expression for the signal-to-noise
ratio from Gaudi \& Winn (2007),
\begin{eqnarray}
S/N &\simeq &0.6 \, \bigg(\frac{m_P}{M_\oplus}\bigg) \bigg(\frac{N_{v}}{100}\bigg)^{1/2} \bigg(\frac{a}{\textrm{AU}}\bigg)^{-1/2} \nonumber \\ 
& & \times \bigg(\frac{m_S}{M_\odot}\bigg)^{-1/2} 10^{- 0.2\,(V - 12)}\frac{D}{3.6\, \textrm{m}},
\label{equn:snr}
\end{eqnarray} 
which was intended to approximate the case of the HARPS instrument on
the ESO 3.6~m telescope. For the challenging case D we assume
$N_v=70$, corresponding to a 20\% uncertainty in the planetary
mass. We perform a MCMC analysis on the simulated data for each
system, just as was done for the 25 transiting systems in this study,
and obtain the corresponding constraints on Trojan companion masses.

Fig.~\ref{fig:kep} shows the simulated data and the results for the
four cases. We find the 2$\sigma$ upper limits on the mass of Trojan
companions to be 25~$M_\oplus$, 26~$M_\oplus$, 21~$M_\oplus$, and
24~$M_\oplus$, for cases A, B, C and D respectively. The results are limited 
by the RV follow-up program; the superb photometric precision of {\it Kepler}\,
does not lead to correspondingly superb constraints on Trojan
masses. 

  The results for cases A--D are all of the same order
  of magnitude. This is because Trojan detectability depends primarily
  on $\sigma_v/\sqrt{N_v}$, which only varies by a factor of 2.2
  between case A and D. For small $m_T/m_P$ and large $\xi$ (see
  Eq.~9), Trojan detectability is indeed independent of planet mass
  for fixed $\sigma_v/\sqrt{N_v}$. A larger planet produces a larger
  RV semi-amplitude, and hence offers greater sensitivity in measuring
  $\Delta t$, but the conversion from $\Delta t$ to $m_T$ varies
  inversely with planetary mass.

  However, for cases C and D, this scaling is not
  precisely obeyed. One reason is that for the larger values of
  $m_T/m_P$ that are relevant in those cases, the relation between
  $m_T$ and $\Delta t$ is nonlinear (see Eq.~\ref{eq:deltat-mt-x}),
  and therefore the error in $m_T$ is not Gaussian. For case D, there
  is an additional source of non-Gaussianity: the signal-to-noise
  ratio is low enough that the correlations between $\Delta t$, $K$,
  $e$, and $\omega$ become important. This means that the upper limit
  on $m_T$ is less constraining than one would predict based only on
  $\sigma_v/\sqrt{N_v}$.

We conclude that a ``serendipitous'' search for Trojan companions to
the habitable-zone planets that will be detected and confirmed as part
of the \textit{Kepler} program will be sensitive to planets of
approximately Neptunian mass or larger. Of course the sensitivity
could be improved by obtaining additional RV data as part of a more
focused search effort. 

Considering the HARPS spectrograph (Pepe et al. 2002), mounted on 
the 3.6 m ESO telescope, as a fiducial instrument for precise
RV measurements, and assuming the noise to be limited by photon-counting statistics,
the measurement uncertainty can be obtained by scaling current
results (Lovis et al.~2005, Gaudi \& Winn, 2007):
\begin{equation}
\sigma_{v} = \frac{10^{0.2\,(V - 12)}}{D/3.6\,\textrm{m}} \textrm{m/s},
\end{equation}
where $V$ is the apparent visual magnitude of the star, and $D$ is the
aperture of the telescope. Here we have assumed a 60~min exposure and
a G-type star.

For small Trojan-to-planet mass ratios, and assuming a circular orbit,
one can determine a nominal estimate of the sensitivity of current
observational facilities to detect Trojan companions to
\textit{Kepler} planets. Under these asumptions, the
signal-to-noise ratio $\Delta t/\sigma_{\Delta t}$ for detecting a
Trojan companion is given by:
\begin{eqnarray}
S/N &\simeq &0.56 \, \bigg(\frac{m_T}{M_\oplus}\bigg) \bigg(\frac{N_{v}}{100}\bigg)^{1/2} \bigg(\frac{a}{\textrm{AU}}\bigg)^{-1/2} \nonumber \\ 
& & \times \bigg(\frac{m_S}{M_\odot}\bigg)^{-1/2} 10^{- 0.2\,(V - 12)}\frac{D}{3.6\, \textrm{m}}
\label{equn:snr}
\end{eqnarray} 
Let us assume a \textit{Kepler} detection of a Jupiter-mass transiting
planet at 1 AU around a sun-like star with $V = 12$, observed with the
HARPS instrument. Then, considering 100 observations evenly spaced in
orbital phase, a 3$\sigma$ Trojan detection (i.e. $\Delta t > 3 \,
\sigma_{\Delta t}$) can be made, if a Trojan mass imbalance of $m_T
\gtrsim 5.36 M_\earth$ existed in the orbit. If the planet were
orbiting instead at 0.03~AU (a ``hot Jupiter'' orbit), then a
detection would be possible for $m_T \gtrsim 0.93 M_\earth$. It would
seem that searching for Trojan companions is a promising alternate
channel for finding small and potentially habitable bodies in the
habitable zones of their parent stars, even if the transiting planets
themselves are too massive to be habitable.

\acknowledgements We thank Jack Wisdom, Scott Gaudi and Eric Ford for
helpful conversations. We further thank Scott Gaudi for providing a
detailed and helpful review of the manuscript. We are grateful to the
William S.~Edgerly Innovation Fund for partial support of this work.

\end{document}